\documentclass[%
reprint,
 amsmath,amssymb,
 aps,
]{revtex4-1}

\usepackage{graphicx}
\usepackage{dcolumn}
\usepackage{bm}
\usepackage{mathtools}
\usepackage{xcolor}

\begin{document}

\preprint{version 2}

\title{Learning to flock through reinforcement}

\author{Mihir Durve}
 \affiliation{Department of Physics, Universit\`a degli studi di 
Trieste, Trieste, Italy 34127}
\affiliation {Quantitative Life Sciences Unit, The Abdus Salam International 
Centre for Theoretical Physics - ICTP, Trieste, Italy 34151}
\author{Fernando Peruani}%
\affiliation{Laboratoire J. A. Dieudonn{\'e}, Universit{\'e} C{\^o}te d'Azur, UMR 7351 CNRS, Parc Valrose, F-06108 Nice Cedex 02, France}
\author{Antonio Celani}
\affiliation{Quantitative Life Sciences Unit, The Abdus Salam International 
Centre for Theoretical Physics - ICTP, Trieste, Italy 34151}
\email{celani@ictp.it}

\date{\today}

\begin{abstract}
Flocks of birds, schools of fish, 
insects swarms are examples of coordinated motion of a group that arises spontaneously from the action of many individuals. 
Here, we study flocking behavior from the viewpoint of multi-agent reinforcement learning. 
In this setting, a learning agent tries to keep contact with the group using as sensory input  the velocity of its neighbors. 
This goal is pursued by each learning individual by exerting  a limited control on its own direction of motion. 
By means of standard reinforcement learning algorithms we show that: 
i) a learning agent exposed to a group of teachers, i.e. hard-wired flocking agents, learns to follow them, 
and ii) that in the absence of teachers, a group of independently learning agents evolves towards 
a state where each agent knows how to flock.  
In both scenarios, i) and ii), the emergent policy (or navigation strategy) corresponds to the polar velocity 
alignment mechanism of the well-known Vicsek model. 
These results show that a) such a velocity alignment may have naturally evolved as an adaptive behavior that aims 
at minimizing the rate of neighbor loss, and b) prove that this alignment does not only favor (local) polar order, but it corresponds to best policy/strategy to keep group cohesion 
when the sensory input is limited to the velocity of neighboring agents. In short, to stay together, steer together.
\end{abstract}

\maketitle


The spectacular collective behavior observed in insect swarms, birds flocks,  and ungulate herds  
 have long fascinated and inspired researchers~\cite{zafeiris,buhl,parrish,ballerini,puckett,ginelli2015intermittent}. There are many long-standing and challenging 
questions about collective animal behavior: How do so many animals 
achieve such a remarkably synchronized motion? What do they perceive 
from their environment and how do they use and share this information within 
the group in order to coordinate their motion? Are there any general rules of 
motion that individuals obey while exhibiting collective behavior? Since the 
last few decades, these questions have been addressed with systematic field observations coupled with mathematical models of animal behavior. Data from experimental observations have been analyzed in order to infer 
the rules that individuals follow in a group~\cite{lukeman,cavagna_MMMAS_2010,herbert,katz,bialek} and 
numerous models have been proposed to explain the observed flocking behavior 
~\cite{aoki,vicsek,couzin,hildenbrandt,cavagna,pearce2014role,barberis}. 
%
%
The great majority of flocking studies~\cite{zafeiris,buhl,parrish,ballerini,puckett,ginelli2015intermittent,lukeman,cavagna_MMMAS_2010,herbert,katz,bialek,aoki,vicsek,couzin,hildenbrandt,cavagna,pearce2014role}, except for few exceptions (see~\cite{barberis} references therein), 
are based on a velocity alignment mechanism that ensures that neighboring individuals move in the same direction. 
However, the origin of such a velocity alignment, from a cognitive point of view, is not known, and neither its biological function.

The natural mathematical language that we will use here to discuss collective motion is the framework of Multi Agent Reinforcement Learning (MARL) ~\cite{panait,busoniu}. In this scheme, the agents can perform actions in response to external cues that they can sense from the environment as well as from other agents. The goal of each agent is to achieve a given objective. In the case at hand, the agents are individuals who can observe the behavior of their close neighbors and react by steering according to some rule.  
 Since it has been hypothesized that there exist many benefits associated to group-living, such as predator avoidance~\cite{milinski} and collective foraging~\cite{pitcher}, 
we assume that the objective of the agents is to increase or maintain the cohesion of the group. The essence of Reinforcement Learning (RL) is that, by repeated trial and error, the agents can learn how to behave in an approximately optimal way so as to achieve their goals~\cite{sutton}. 
Here, we show that velocity alignment emerges spontaneously in a RL process from the minimization of the rate of neighbor loss, 
and represents a optimal strategy to keep group cohesion.

In the following we will consider individual agents that move at constant speed in a two-dimensional box with periodic boundary conditions. The density of agents is kept fixed to $\rho=2$ agents/(unit length)$^2$.
Updates are performed at discrete time steps as follows. For the i-th agent, the position update is: 
 \begin{equation}
 \label{p_u}
  \mathbf{r}^{t+1}_i = \mathbf{r}^{t}_i + v_0 \mathbf{v}^{t}_i \Delta t \; ,
\end{equation}
where $\mathbf{r}^{t}_i$  and $\mathbf{v}^{t}_i$, with $||\mathbf{v}^{t}_i||=1$, are the position and moving direction, respectively, of the agent  at time $t$; 
the term $v_0$ corresponds to the speed, which we fix to $v_0=0.5$, and $\Delta t =1$.  
At each time step, each agents makes a decision on whether keeping the current heading direction or performing a turn. 
The decision-making process is based on the sensorial input of the agent, which corresponds to the angular difference between the (normalized) average 
velocity defined by $\mathbf{P}_i = (\sum_{|\mathbf{r}_j - \mathbf{r}_i| < R} \mathbf{v}^{t}_j)/n_i$  (with $n_i$ the number of neighbors of agent $i$ within its perception range $R$) and  the moving direction of the agent $\mathbf{v}^{t}_i$.  Below, we take $R=1$.
We can express the state as 
\begin{equation} s^t_ i = \arg(\mathbf{P}_i,  \mathbf{v}^t_i) \, ,
 \label{relative-angle}
\end{equation}
where the function $\arg(\mathbf{P}_i,  \mathbf{v}^t_i)$ is defined as $\arccos(\mathbf{P}_i \cdot  \mathbf{v}^t_i/||\mathbf{P}_i||)$ for  
$\mathbf{P}_i \cdot  {(\mathbf{v}^t_i)}_{\perp}>0$ with ${(\mathbf{v}^t_i)}_{\perp}$ obtained by rotating $\pi/2$ counter clockwise the unit vector $\mathbf{v}^t_i$, and minus this quantity otherwise. 
This means that $s^t_i \in [-\pi,\pi)$. 
For computational simplicity, we discretize $s^t_i$ by dividing $2 \pi$ into $K_s$ equally spaced elements.
%
In the RL language, the relative angle $s^t_i$ is the contextual information that defines the current state of the i-th agent. 
Knowing $s^t_i$, the agent updates $\mathbf{v}^t_i$ by turning this vector an angle $a^t_i$ 
\begin{equation}
\label{v_u}
 \mathbf{v}_{t+1} = \mathcal{R}(a^t_i) \mathbf{v}_{t} \; ,
 \end{equation}
where $\mathcal{R}(a^t_i)$ is a rotation matrix. 
Note that there are  $K_a$ possible turning angles, equally spaced in $[-\theta_{\mathit{max}},\theta_{\mathit{max}}]$.
In the RL jargon, choosing the turning angle $a^t_i$ represents an "action" performed by the agent. 
The association of a given state $s^i_t$ with an action  $a^i_t$ is called a policy.  
Policy evaluation takes place at each time step as the agent receives a (negative) reinforcement signal in the form of a cost $c^{t+1}_i$ for losing neighbors within its perception range $R$
\begin{equation}
\label{cost}
c^{t+1}_i= 
        \left\{ \begin{array}{ll}
	1, & \qquad \text{if } n^{t+1}_i < n^t_i\\
	0, & \qquad \text{otherwise}
        \end{array} \right.
    \end{equation}
where $n^t_i$ is the current number of neighbors. 
The goal of the learning agent is to find a policy that minimizes the cost. 
To achieve this goal the agent makes use of a simple learning rule \cite{watkins,sutton}. 
The i-th learning agent keeps in memory a table of values $Q_i(s,a)$ for each state-action pair (here a matrix $K_a\times K_s$) which is updated at each step of the dynamics -- only for the entry that corresponds to the state just visited and the action just taken -- according to  
\begin{equation}
\label{q-update}
 Q_i(s^t_i,a^t_i) \leftarrow Q_i(s^t_i,a^t_i) + \alpha[c^{t+1}_i - Q_i(s^t_i,a^t_i) ] \, .
\end{equation}
This update rule effectively constructs an estimator for the expected cost that will be incurred by starting in a given state and performing a given action.
The policy at each time-step is based upon the current $Q_i$ according to so-called $\epsilon-$greedy exploration scheme:
\begin{equation}
\label{epsilon-greedy}
a^{t}_i= 
        \left\{ \begin{array}{ll}
	\underset{a'}{\operatorname{argmin}} \hspace{2mm} Q_i(s^t_i,a') & \text{with 
prob. } \hspace{2mm} 1-\epsilon\\
	\text{an action at random} & \text{with prob.} \hspace{2mm} 
\epsilon
        \end{array} \right. \; .
    \end{equation} 
In the simulations we have used $\alpha=0.005$ and various different schedules for the exploration probability \cite{sutton}.

\begin{figure} [!t]
\includegraphics[width=0.5\textwidth]{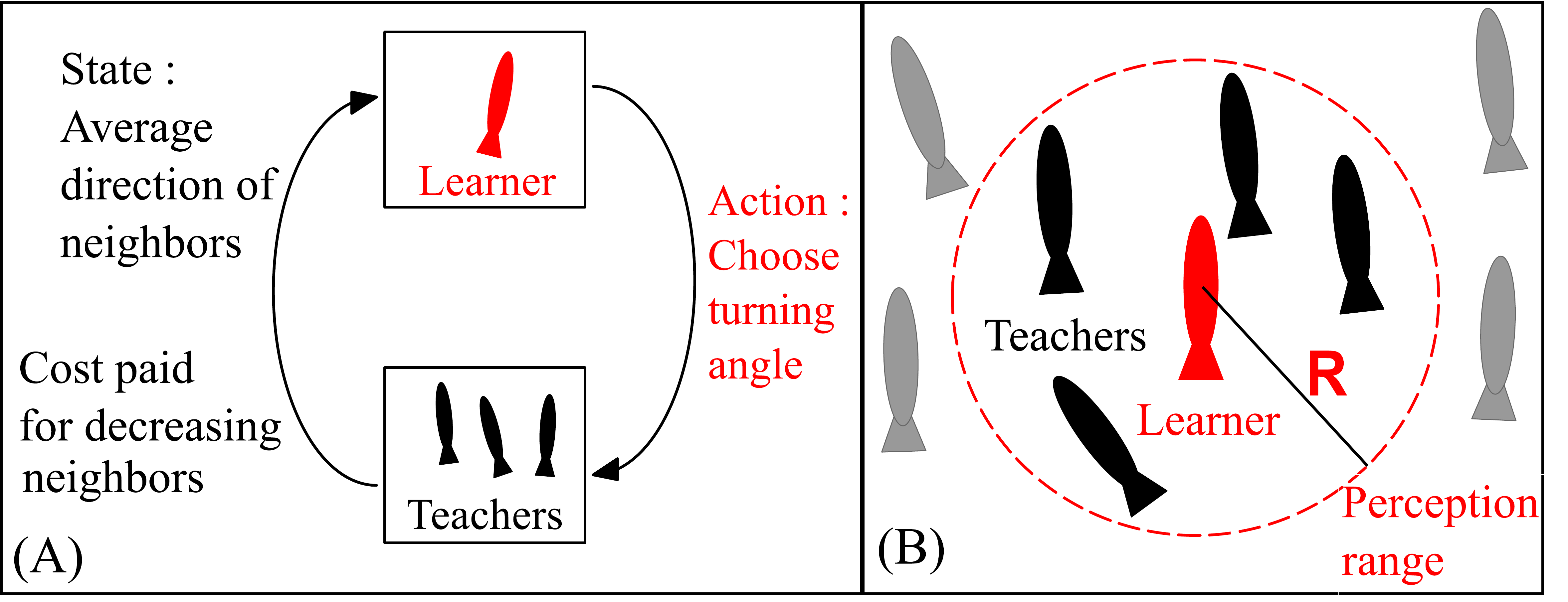}
\caption{\label{neighborhood} Learning to flock within a group of teachers. (A) Scheme of reinforcement learning. (B) 
Neighbors (black) within the perception range of the learner (red). }
\end{figure}

We start by considering the case when there is a single learning agent in a crowd of $N$ teachers (see Figure~\ref{neighborhood}) who have a hard-wired 
policy:
\begin{equation}
\label{policy}
a^t_i(s^t_i)= 
        \left\{ \begin{array}{ll}
	s^t_i & \text{if } |s^t_i| \leq \theta_{\mathit{max}}\\
	 \theta_{\mathit{max}} &\text{if } s^t_i >  
\theta_{\mathit{max}}\\
-\theta_{\mathit{max}} &\text{if } s^t_i <  
- \theta_{\mathit{max}}\\
        \end{array} \right. \;.
    \end{equation}
This decision rule is nothing else but a version of the Vicsek model of flocking with a discrete number of possible moving direction and limited angular speed. 
Thus, teachers display robust collective motion. 
The learning agent, on the contrary, does not have a fixed policy, but one that evolves over time as it acquires experience, i.e. by visiting a large number states, 
and evaluating the outcome of executing its actions.  
In this case we find that  for suitably chosen learning rates $\alpha$ and exploration probability $\epsilon$ the algorithm approaches an approximately optimal solution to the decision-making problem after some  period of training. 

\begin{figure*} [!ht]
\includegraphics[width=0.8\textwidth]{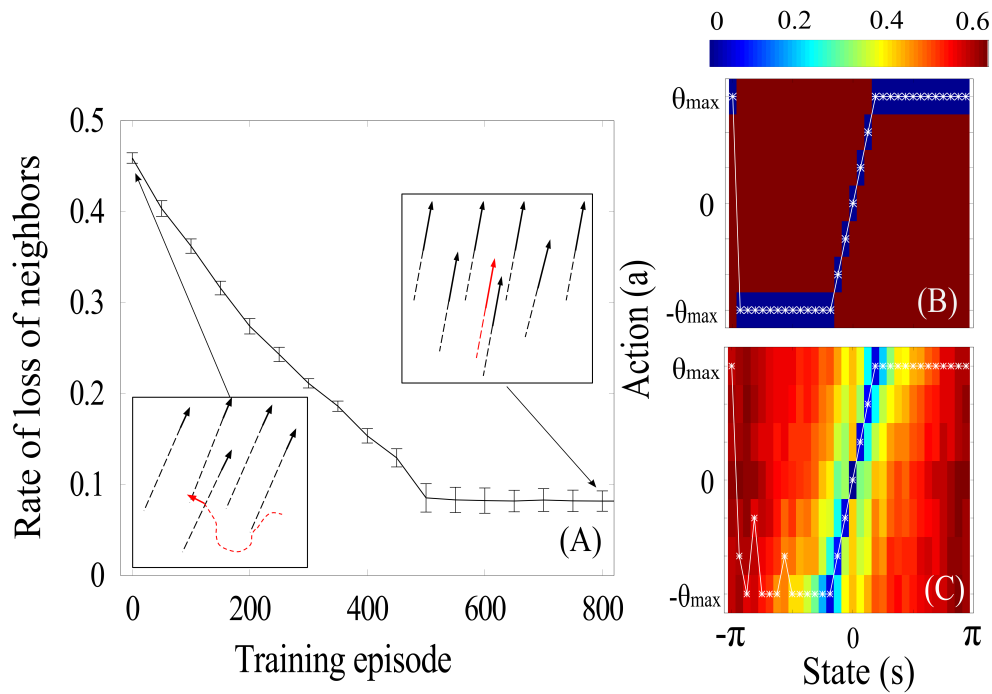}
\caption{Single learner results. (A) Performance of the learner as training progresses. 
The error bars indicate standard deviation in the values in 20 training 
sessions. In the inset, some short trajectories of the learning agent (red) and teachers 
(black) at the stages indicated by the arrows. The number of teachers is $N=200$. The maximal turning angle is $\theta_{\mathit{max}}=3\pi/16$.  (B) The Q-matrix, i.e the average cost incurred for a given state-action pair by 
the teachers. White stars shows the action $a$ taken by the teacher when in state 
$s$. (C) Q-matrix of the learner at the end the training session. White stars denote the best estimated action of the learner for each state.} \label{single} 
\end{figure*}  

In  simulations, we break the training session into a number of
training episodes of equal prescribed duration of $10^4$ time steps. In each 
episode the teachers start with random initial positions and velocities.
After a transient, they form an ordered flock and at this time we introduce the learner and implement the learning algorithm. At the very beginning, the 
learner starts with a Q-matrix with all zero entries, which in a case of optimistic initialization (the naive learner expects to incur no costs), a choice that is known to favor exploration~\cite{sutton}. From one episode to the following, the learner keeps in memory the Q-matrix that it has learned so far. 
During the training session, we measure the success of the learning process with 
the average cost that a learner 
pays per time step, that is the rate at which it is losing contact with the teachers (see Fig.~\ref{single}A).
As the training progresses, the rate at which neighbors are lost starts from an initial value of $0.5$, meaning that on average the learner loses contact with some neighbor every other step, to decrease and eventually saturate down to a value around $0.1$ meaning that the contact is kept for 90\% of the time. In the insets of  
Fig~\ref{single}A we show samples of short trajectories of the learner and some teachers at the 
early and later phase of the training process. We 
observe that in the early phase of the training, the 
learner essentially moves at random (see movie1.mp4) and eventually it learns to stay within the flock (see 
movie2.mp4).
In Fig~\ref{single}C we show that  
the policy discovered by the learner is  identical with the pre-defined 
policy of the teachers, see Eq. (\ref{policy}) and Fig.~\ref{single}B. It is important to remark that the one and only goal of the learner is to keep contact with its neighboring teachers, not to imitate their behavior. It simply turns out that the best strategy for the learner is in fact the
underlying rule that was assigned to teachers.

\begin{figure} [!b]
\includegraphics[width=0.5\textwidth]{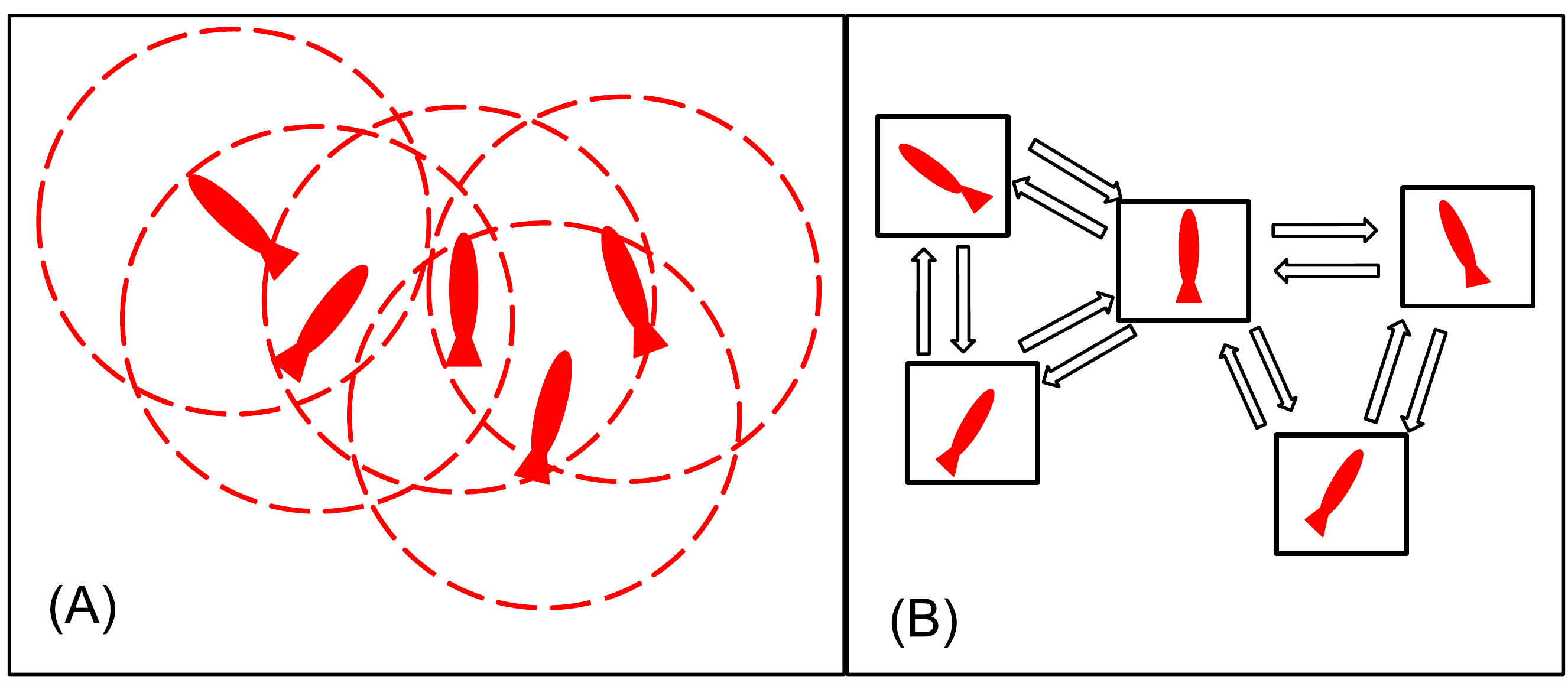}
\caption{\label{Nice_Fig3} Multi-agent concurrent learning. (A) Multiple agents with their perception range. 
(B) Agents interact with short-range, reciprocal interactions.}
\end{figure}

 Now, let us move our focus to the situation where there are no teachers, but only $N$ independently learning agents (see Figure~\ref{Nice_Fig3}).
A distinctive difficulty of applying reinforcement learning to the multi-agent setting is that all individuals have to concurrently discover the best way to behave and cannot rely upon knowledge previously acquired by their peers. 
However, we find  that $N$ learning agents are able to overcome this hurdle and are actually capable of learning to flock even in the case when all of them start as absolute beginners (all Q-matrices initialized to zero). 

\begin{figure*} [!ht]
\includegraphics[width=0.8\textwidth]{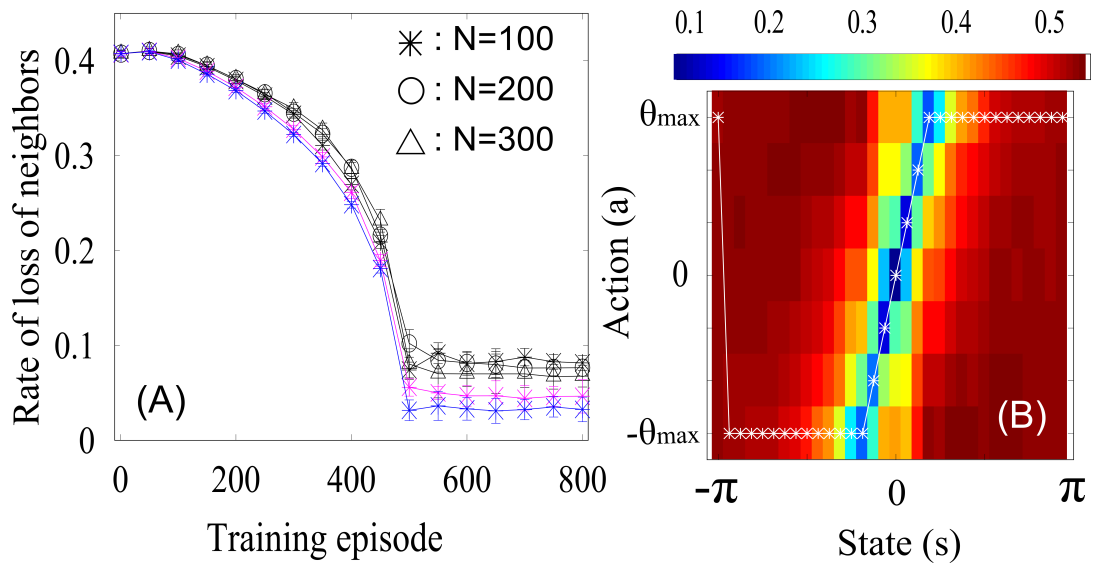}
\caption{\label{m_re} Results for the multi-agent concurrent learning. (A) Average performance of learners in groups of different sizes. Black, magenta and blue 
colors corresponds to state-action spaces of size \{$Ks,Ka$\}=\{32, 7\}, \{64, 14\}, 
\{128, 28\} respectively. Error bars indicate standard deviation in the 
average values for each agent. (B) Average Q-matrix at the end of the training for 
$N=200$ agents with combination of \{$Ks,Ka$\}=\{32, 7\}. White points indicate 
actions with estimated minimum cost for given state. The colors represent values 
in the Q-matrix.}
\end{figure*}

To characterize the performance of the learners, we measure the average rate of loss of neighbors. In Fig~\ref{m_re}A we show the average cost for various groups sizes and 
state-action space discretizations $\{K_s,K_a\}$. The cost reaches a small and steady value after few hundreds of episodes. As the group size grows, the performance remains essentially the same. Conversely, refining the discretization allows to further reduce the costs: for 128 relative alignment angles and 28 turning angles the agents  do not lose neighbors for about 97\% of the time.

The resulting Q-matrix at the end of the training, averaged over all learners, is shown in Fig.~\ref{m_re}B. The colors represent the numerical values in the Q-matrix and 
the  discovered policy is shown with white points.  
We observe that the discovered policy is the same that the one learned by the single agent with $N$ teachers. 

\begin{figure} [!hb]
\includegraphics[width=0.5\textwidth]{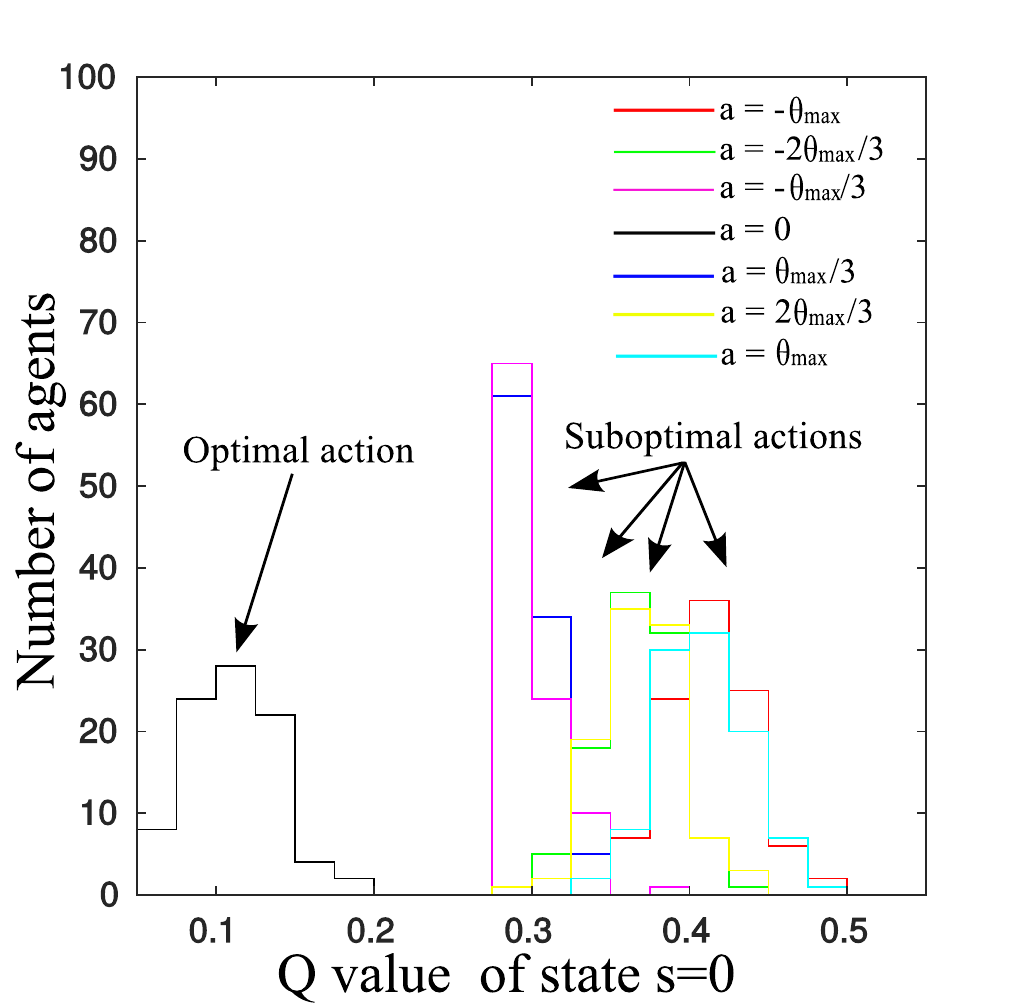}
\caption{\label{fluctuations} All agents independently learn the same optimal strategy. The histogram shows the frequency of a give numerical value of $Q(0,a)$ across all learners, at the end of training. The best action $a=0$ always performs better than any other action. The same holds for other states (not shown). Data obtained with $N=100$ agents, $K_s=32$, $K_a=7$.}
\end{figure}

It is worth stressing that all agents independently learn the same strategy. We have collected the values of the Q-matrix for a given state ($s=0$) and different actions, for all agents, at the end of training. The histogram for the frequency of Q values is shown in Fig.~\ref{fluctuations} where one can observe that there is a clear gap that separates the estimated costs for the optimal turning angle, which lie around 0.1, from  the suboptimal actions that have significantly larger costs.

\begin{figure} [!ht]
\includegraphics[width=0.5\textwidth]{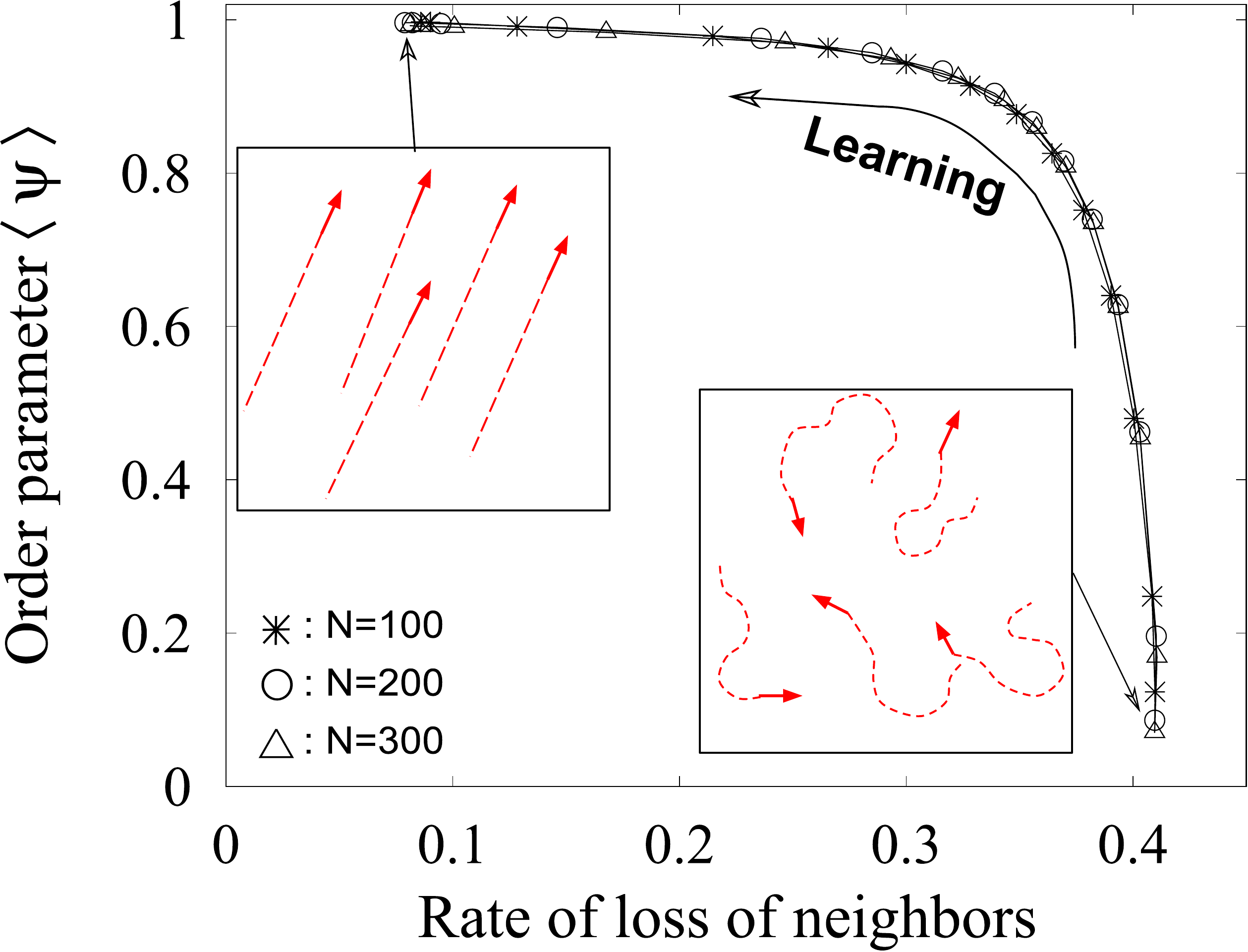}
\caption{\label{m_op} Average polar order parameter $\langle \psi \rangle$ versus the rate
of loss of neighbors. In the insets we show a few short trajectories 
of naive and trained agents. }
\end{figure}

A customary measure of the degree of alignment
is the polar order parameter:
\begin{equation}
 \psi(t) = \frac{1}{N} \hspace{1mm}  \hspace{1mm}\left \vert \left\vert \sum_{i=1}^{N} 
\mathbf{v}^t_{i}\hspace{1mm}\right \vert\right\vert. \label{eq:orpar}
 \end{equation} 
 If all the agents are oriented randomly then, as $N \rightarrow \infty$, $\psi \rightarrow 0$ whereas if all the agents are oriented in the same 
direction then $\psi=1$.
In Fig~\ref{m_op} we show the evolution of order parameter versus the average cost as the multi-agent learning is advancing. We 
observe that in the early phases of training the rate of loss of neighbors 
is comparatively high and the direction consensus among the agents is low, in agreement with the notion that the 
agents are behaving randomly (see movie3.mp4). As the learning 
progresses, the agents discover how to keep cohesion, and in doing so they achieve a highly ordered state (see 
movie4.mp4). 

Therefore we conclude that the obtained results  
proves that the velocity alignment mechanism of the Vicsek model (see Eq. (\ref{policy})) -- based on energy minimization of spin-spin interaction of the XY model -- 
can spontaneously emerge, counterintuitively, from the minimization of neighbor-loss rate, 
and furthermore represents an optimal strategy to keep group cohesion when the perception is limited to velocity of neighboring agents.
In summary, if the agents want to stay together, they must learn that they have to steer together.

In more general terms, we have shown that Multi-Agent Reinforcement Learning can provide an effective way to deal with questions about the emergence of collective behaviors and the driving forces behind them. Our present contribution is just an initial step in this direction and we feel that prospective applications of this approach remain largely unexplored. 

For instance, in the present work we have decided at the outset the structure of the perceptual space of the agents, namely the choice of the radius of perception as the relevant parameter and of the relative angle as the relevant state variable. In doing so, we bypassed fundamental questions like: Is the metric distance the most appropriate choice for ranking neighbors~? How should the information given by other individuals be discounted depending on their ranking~? A more ambitious approach would tackle these issues directly through MARL and try to learn from experience what are better choices of the state variable that allow to achieve optimal cohesion. 

As another example, here we have tasked our agents with the goal of keeping contact with neighbors, which in itself is understood to be a secondary goal motivated by the primary need of avoiding predators (safety by the numbers) or of increasing the efficiency of foraging. Can one recapitulate the congregation behavior by tasking agents with the primary goal itself~? More explicitly, would agents learn to align themselves by rewarding behaviors that reduce their risk of being predated or increase their chance of getting some food~? 

Also, in this paper we have considered a group of identical agents. When agents differ for their perceptual abilities or their dexterity in taking the appropriate actions, then competitive behaviors may arise within the group and the problem acquires a new challenging dimension. How much heterogeneity and competition can be tolerated before it
starts impacting the benefit of staying in a group~?

These and many other questions lend themselves to be attacked by the techniques of MARL and we believe that the approach that we have delineated here will show its full potential in the near future.

M.~D. acknowledges fruitful discussion with A. Pezzotta, M. Adorisio and A. Mazzolini.
M.~D. is grateful for the graduate fellowship by the ICTP 
and University of Trieste. M.~D. and A.~C. acknowledge hospitality and support from 
Laboratoire J. A. Dieudonn{\'e}, Universit{\'e} C{\^o}te d'Azur, France.

\bibliography{main}

\begin{thebibliography}{24}%
\makeatletter
\providecommand \@ifxundefined [1]{%
 \@ifx{#1\undefined}
}%
\providecommand \@ifnum [1]{%
 \ifnum #1\expandafter \@firstoftwo
 \else \expandafter \@secondoftwo
 \fi
}%
\providecommand \@ifx [1]{%
 \ifx #1\expandafter \@firstoftwo
 \else \expandafter \@secondoftwo
 \fi
}%
\providecommand \natexlab [1]{#1}%
\providecommand \enquote  [1]{``#1''}%
\providecommand \bibnamefont  [1]{#1}%
\providecommand \bibfnamefont [1]{#1}%
\providecommand \citenamefont [1]{#1}%
\providecommand \href@noop [0]{\@secondoftwo}%
\providecommand \href [0]{\begingroup \@sanitize@url \@href}%
\providecommand \@href[1]{\@@startlink{#1}\@@href}%
\providecommand \@@href[1]{\endgroup#1\@@endlink}%
\providecommand \@sanitize@url [0]{\catcode `\\12\catcode `\$12\catcode
  `\&12\catcode `\#12\catcode `\^12\catcode `\_12\catcode `\%12\relax}%
\providecommand \@@startlink[1]{}%
\providecommand \@@endlink[0]{}%
\providecommand \url  [0]{\begingroup\@sanitize@url \@url }%
\providecommand \@url [1]{\endgroup\@href {#1}{\urlprefix }}%
\providecommand \urlprefix  [0]{URL }%
\providecommand \Eprint [0]{\href }%
\providecommand \doibase [0]{http://dx.doi.org/}%
\providecommand \selectlanguage [0]{\@gobble}%
\providecommand \bibinfo  [0]{\@secondoftwo}%
\providecommand \bibfield  [0]{\@secondoftwo}%
\providecommand \translation [1]{[#1]}%
\providecommand \BibitemOpen [0]{}%
\providecommand \bibitemStop [0]{}%
\providecommand \bibitemNoStop [0]{.\EOS\space}%
\providecommand \EOS [0]{\spacefactor3000\relax}%
\providecommand \BibitemShut  [1]{\csname bibitem#1\endcsname}%
\let\auto@bib@innerbib\@empty
\bibitem [{\citenamefont {Vicsek}\ and\ \citenamefont
  {Zafeiris}(2012)}]{zafeiris}%
  \BibitemOpen
  \bibfield  {author} {\bibinfo {author} {\bibfnamefont {T.}~\bibnamefont
  {Vicsek}}\ and\ \bibinfo {author} {\bibfnamefont {A.}~\bibnamefont
  {Zafeiris}},\ }\href@noop {} {\bibfield  {journal} {\bibinfo  {journal}
  {Phys. Rep.}\ }\textbf {\bibinfo {volume} {517}},\ \bibinfo {pages} {71}
  (\bibinfo {year} {2012})}\BibitemShut {NoStop}%
\bibitem [{\citenamefont {Buhl}\ \emph {et~al.}(2006)\citenamefont {Buhl},
  \citenamefont {Sumpter}, \citenamefont {Couzin}, , \citenamefont {Hale},
  \citenamefont {Despland}, \citenamefont {Miller},\ and\ \citenamefont
  {Simpson.}}]{buhl}%
  \BibitemOpen
  \bibfield  {author} {\bibinfo {author} {\bibfnamefont {J.}~\bibnamefont
  {Buhl}}, \bibinfo {author} {\bibfnamefont {D.}~\bibnamefont {Sumpter}},
  \bibinfo {author} {\bibfnamefont {I.}~\bibnamefont {Couzin}}, , \bibinfo
  {author} {\bibfnamefont {J.}~\bibnamefont {Hale}}, \bibinfo {author}
  {\bibfnamefont {E.}~\bibnamefont {Despland}}, \bibinfo {author}
  {\bibfnamefont {E.}~\bibnamefont {Miller}}, \ and\ \bibinfo {author}
  {\bibfnamefont {S.}~\bibnamefont {Simpson.}},\ }\href@noop {} {\bibfield
  {journal} {\bibinfo  {journal} {Science}\ }\textbf {\bibinfo {volume}
  {312}},\ \bibinfo {pages} {1402} (\bibinfo {year} {2006})}\BibitemShut
  {NoStop}%
\bibitem [{\citenamefont {Parrish}\ \emph {et~al.}(2002)\citenamefont
  {Parrish}, \citenamefont {Viscido},\ and\ \citenamefont
  {Gr{\"unbaum}}}]{parrish}%
  \BibitemOpen
  \bibfield  {author} {\bibinfo {author} {\bibfnamefont {J.}~\bibnamefont
  {Parrish}}, \bibinfo {author} {\bibfnamefont {S.}~\bibnamefont {Viscido}}, \
  and\ \bibinfo {author} {\bibfnamefont {D.}~\bibnamefont {Gr{\"unbaum}}},\
  }\href@noop {} {\bibfield  {journal} {\bibinfo  {journal} {Biol. Bull}\
  }\textbf {\bibinfo {volume} {202}},\ \bibinfo {pages} {296} (\bibinfo {year}
  {2002})}\BibitemShut {NoStop}%
\bibitem [{\citenamefont {Ballerini}\ \emph {et~al.}(2008)\citenamefont
  {Ballerini}, \citenamefont {Cabibbo}, \citenamefont {Candelier},
  \citenamefont {Cavagna}, \citenamefont {Cisbani}, \citenamefont {Giardina},
  \citenamefont {Lecomte}, \citenamefont {Orlandi}, \citenamefont {Parisi},
  \citenamefont {Procaccini}, \citenamefont {Viale},\ and\ \citenamefont
  {Zdravkovic}}]{ballerini}%
  \BibitemOpen
  \bibfield  {author} {\bibinfo {author} {\bibfnamefont {M.}~\bibnamefont
  {Ballerini}}, \bibinfo {author} {\bibfnamefont {N.}~\bibnamefont {Cabibbo}},
  \bibinfo {author} {\bibfnamefont {R.}~\bibnamefont {Candelier}}, \bibinfo
  {author} {\bibfnamefont {A.}~\bibnamefont {Cavagna}}, \bibinfo {author}
  {\bibfnamefont {E.}~\bibnamefont {Cisbani}}, \bibinfo {author} {\bibfnamefont
  {I.}~\bibnamefont {Giardina}}, \bibinfo {author} {\bibfnamefont
  {V.}~\bibnamefont {Lecomte}}, \bibinfo {author} {\bibfnamefont
  {A.}~\bibnamefont {Orlandi}}, \bibinfo {author} {\bibfnamefont
  {G.}~\bibnamefont {Parisi}}, \bibinfo {author} {\bibfnamefont
  {A.}~\bibnamefont {Procaccini}}, \bibinfo {author} {\bibfnamefont
  {M.}~\bibnamefont {Viale}}, \ and\ \bibinfo {author} {\bibfnamefont
  {V.}~\bibnamefont {Zdravkovic}},\ }\href@noop {} {\bibfield  {journal}
  {\bibinfo  {journal} {PNAS}\ }\textbf {\bibinfo {volume} {105}},\ \bibinfo
  {pages} {1232} (\bibinfo {year} {2008})}\BibitemShut {NoStop}%
\bibitem [{\citenamefont {Puckett}\ \emph {et~al.}(2014)\citenamefont
  {Puckett}, \citenamefont {Kelley},\ and\ \citenamefont
  {Ouellette}}]{puckett}%
  \BibitemOpen
  \bibfield  {author} {\bibinfo {author} {\bibfnamefont {J.}~\bibnamefont
  {Puckett}}, \bibinfo {author} {\bibfnamefont {D.}~\bibnamefont {Kelley}}, \
  and\ \bibinfo {author} {\bibfnamefont {N.}~\bibnamefont {Ouellette}},\
  }\href@noop {} {\bibfield  {journal} {\bibinfo  {journal} {Scientific
  Reports}\ }\textbf {\bibinfo {volume} {4}} (\bibinfo {year}
  {2014})}\BibitemShut {NoStop}%
\bibitem [{\citenamefont {Ginelli}\ \emph {et~al.}(2015)\citenamefont
  {Ginelli}, \citenamefont {Peruani}, \citenamefont {Pillot}, \citenamefont
  {Chat{\'e}}, \citenamefont {Theraulaz},\ and\ \citenamefont
  {Bon}}]{ginelli2015intermittent}%
  \BibitemOpen
  \bibfield  {author} {\bibinfo {author} {\bibfnamefont {F.}~\bibnamefont
  {Ginelli}}, \bibinfo {author} {\bibfnamefont {F.}~\bibnamefont {Peruani}},
  \bibinfo {author} {\bibfnamefont {M.-H.}\ \bibnamefont {Pillot}}, \bibinfo
  {author} {\bibfnamefont {H.}~\bibnamefont {Chat{\'e}}}, \bibinfo {author}
  {\bibfnamefont {G.}~\bibnamefont {Theraulaz}}, \ and\ \bibinfo {author}
  {\bibfnamefont {R.}~\bibnamefont {Bon}},\ }\href@noop {} {\bibfield
  {journal} {\bibinfo  {journal} {Proceedings of the National Academy of
  Sciences}\ }\textbf {\bibinfo {volume} {112}},\ \bibinfo {pages} {12729}
  (\bibinfo {year} {2015})}\BibitemShut {NoStop}%
\bibitem [{\citenamefont {Lukeman}\ \emph {et~al.}(2010)\citenamefont
  {Lukeman}, \citenamefont {Li},\ and\ \citenamefont
  {Edelstein-Keshet}}]{lukeman}%
  \BibitemOpen
  \bibfield  {author} {\bibinfo {author} {\bibfnamefont {R.}~\bibnamefont
  {Lukeman}}, \bibinfo {author} {\bibfnamefont {Y.-X.}\ \bibnamefont {Li}}, \
  and\ \bibinfo {author} {\bibfnamefont {L.}~\bibnamefont {Edelstein-Keshet}},\
  }\href@noop {} {\bibfield  {journal} {\bibinfo  {journal} {Proceedings of the
  National Academy of Sciences}\ }\textbf {\bibinfo {volume} {107}},\ \bibinfo
  {pages} {12576} (\bibinfo {year} {2010})}\BibitemShut {NoStop}%
\bibitem [{\citenamefont {Cavagna}\ \emph {et~al.}(2010)\citenamefont
  {Cavagna}, \citenamefont {Cimarelli}, \citenamefont {Giardina}, \citenamefont
  {Parisi}, \citenamefont {Santagati}, \citenamefont {Stefanini},\ and\
  \citenamefont {Tavarone}}]{cavagna_MMMAS_2010}%
  \BibitemOpen
  \bibfield  {author} {\bibinfo {author} {\bibfnamefont {A.}~\bibnamefont
  {Cavagna}}, \bibinfo {author} {\bibfnamefont {A.}~\bibnamefont {Cimarelli}},
  \bibinfo {author} {\bibfnamefont {I.}~\bibnamefont {Giardina}}, \bibinfo
  {author} {\bibfnamefont {G.}~\bibnamefont {Parisi}}, \bibinfo {author}
  {\bibfnamefont {R.}~\bibnamefont {Santagati}}, \bibinfo {author}
  {\bibfnamefont {F.}~\bibnamefont {Stefanini}}, \ and\ \bibinfo {author}
  {\bibfnamefont {R.}~\bibnamefont {Tavarone}},\ }\href@noop {} {\bibfield
  {journal} {\bibinfo  {journal} {Mathematical Models and Methods in Applied
  Sciences}\ }\textbf {\bibinfo {volume} {20}},\ \bibinfo {pages} {1491}
  (\bibinfo {year} {2010})}\BibitemShut {NoStop}%
\bibitem [{\citenamefont {Herbert-Read}\ \emph {et~al.}(2011)\citenamefont
  {Herbert-Read}, \citenamefont {Perna}, \citenamefont {Mann}, \citenamefont
  {Schaerf}, \citenamefont {Sumpter},\ and\ \citenamefont {Ward}}]{herbert}%
  \BibitemOpen
  \bibfield  {author} {\bibinfo {author} {\bibfnamefont {J.}~\bibnamefont
  {Herbert-Read}}, \bibinfo {author} {\bibfnamefont {A.}~\bibnamefont {Perna}},
  \bibinfo {author} {\bibfnamefont {R.}~\bibnamefont {Mann}}, \bibinfo {author}
  {\bibfnamefont {T.}~\bibnamefont {Schaerf}}, \bibinfo {author} {\bibfnamefont
  {D.}~\bibnamefont {Sumpter}}, \ and\ \bibinfo {author} {\bibfnamefont
  {A.}~\bibnamefont {Ward}},\ }\href@noop {} {\bibfield  {journal} {\bibinfo
  {journal} {Proceedings of the National Academy of Sciences}\ }\textbf
  {\bibinfo {volume} {108}},\ \bibinfo {pages} {18726} (\bibinfo {year}
  {2011})}\BibitemShut {NoStop}%
\bibitem [{\citenamefont {Katz}\ \emph {et~al.}(2011)\citenamefont {Katz},
  \citenamefont {Tunstr{\o}m}, \citenamefont {Ioannou}, \citenamefont {Huepe},\
  and\ \citenamefont {Couzin}}]{katz}%
  \BibitemOpen
  \bibfield  {author} {\bibinfo {author} {\bibfnamefont {Y.}~\bibnamefont
  {Katz}}, \bibinfo {author} {\bibfnamefont {K.}~\bibnamefont {Tunstr{\o}m}},
  \bibinfo {author} {\bibfnamefont {C.}~\bibnamefont {Ioannou}}, \bibinfo
  {author} {\bibfnamefont {C.}~\bibnamefont {Huepe}}, \ and\ \bibinfo {author}
  {\bibfnamefont {I.}~\bibnamefont {Couzin}},\ }\href@noop {} {\bibfield
  {journal} {\bibinfo  {journal} {Proceedings of the National Academy of
  Sciences}\ }\textbf {\bibinfo {volume} {108}},\ \bibinfo {pages} {18720}
  (\bibinfo {year} {2011})}\BibitemShut {NoStop}%
\bibitem [{\citenamefont {Bialek}\ \emph {et~al.}(2012)\citenamefont {Bialek},
  \citenamefont {Cavagna}, \citenamefont {Giardina}, \citenamefont {Mora},
  \citenamefont {Silvestri}, \citenamefont {Viale},\ and\ \citenamefont
  {Walczak}}]{bialek}%
  \BibitemOpen
  \bibfield  {author} {\bibinfo {author} {\bibfnamefont {W.}~\bibnamefont
  {Bialek}}, \bibinfo {author} {\bibfnamefont {A.}~\bibnamefont {Cavagna}},
  \bibinfo {author} {\bibfnamefont {I.}~\bibnamefont {Giardina}}, \bibinfo
  {author} {\bibfnamefont {T.}~\bibnamefont {Mora}}, \bibinfo {author}
  {\bibfnamefont {E.}~\bibnamefont {Silvestri}}, \bibinfo {author}
  {\bibfnamefont {M.}~\bibnamefont {Viale}}, \ and\ \bibinfo {author}
  {\bibfnamefont {A.}~\bibnamefont {Walczak}},\ }\href@noop {} {\bibfield
  {journal} {\bibinfo  {journal} {Proceedings of the National Academy of
  Sciences}\ }\textbf {\bibinfo {volume} {109}},\ \bibinfo {pages} {4786}
  (\bibinfo {year} {2012})}\BibitemShut {NoStop}%
\bibitem [{\citenamefont {Aoki}(1982)}]{aoki}%
  \BibitemOpen
  \bibfield  {author} {\bibinfo {author} {\bibfnamefont {I.}~\bibnamefont
  {Aoki}},\ }\href@noop {} {\bibfield  {journal} {\bibinfo  {journal} {Bulletin
  of the Japanese Society of Scientific Fisheries}\ }\textbf {\bibinfo {volume}
  {48(8)}},\ \bibinfo {pages} {1081} (\bibinfo {year} {1982})}\BibitemShut
  {NoStop}%
\bibitem [{\citenamefont {Vicsek}\ \emph {et~al.}(1995)\citenamefont {Vicsek},
  \citenamefont {Czir{\'o}k}, \citenamefont {Ben-Jacob}, \citenamefont
  {Cohen},\ and\ \citenamefont {Shochet}}]{vicsek}%
  \BibitemOpen
  \bibfield  {author} {\bibinfo {author} {\bibfnamefont {T.}~\bibnamefont
  {Vicsek}}, \bibinfo {author} {\bibfnamefont {A.}~\bibnamefont {Czir{\'o}k}},
  \bibinfo {author} {\bibfnamefont {E.}~\bibnamefont {Ben-Jacob}}, \bibinfo
  {author} {\bibfnamefont {I.}~\bibnamefont {Cohen}}, \ and\ \bibinfo {author}
  {\bibfnamefont {O.}~\bibnamefont {Shochet}},\ }\href@noop {} {\bibfield
  {journal} {\bibinfo  {journal} {Phys.\ Rev.\ Lett}\ }\textbf {\bibinfo
  {volume} {75}},\ \bibinfo {pages} {1226} (\bibinfo {year}
  {1995})}\BibitemShut {NoStop}%
\bibitem [{\citenamefont {Couzin}\ and\ \citenamefont {Krause}(2003)}]{couzin}%
  \BibitemOpen
  \bibfield  {author} {\bibinfo {author} {\bibfnamefont {I.~D.}\ \bibnamefont
  {Couzin}}\ and\ \bibinfo {author} {\bibfnamefont {J.}~\bibnamefont
  {Krause}},\ }\href@noop {} {\bibfield  {journal} {\bibinfo  {journal} {Adv
  Study Behav}\ }\textbf {\bibinfo {volume} {32}},\ \bibinfo {pages} {1}
  (\bibinfo {year} {2003})}\BibitemShut {NoStop}%
\bibitem [{\citenamefont {Hildenbrandt}\ \emph {et~al.}(2010)\citenamefont
  {Hildenbrandt}, \citenamefont {Carere},\ and\ \citenamefont
  {Hemelrijk}}]{hildenbrandt}%
  \BibitemOpen
  \bibfield  {author} {\bibinfo {author} {\bibfnamefont {H.}~\bibnamefont
  {Hildenbrandt}}, \bibinfo {author} {\bibfnamefont {C.}~\bibnamefont
  {Carere}}, \ and\ \bibinfo {author} {\bibfnamefont {C.}~\bibnamefont
  {Hemelrijk}},\ }\href@noop {} {\bibfield  {journal} {\bibinfo  {journal}
  {Behavioral Ecology}\ }\textbf {\bibinfo {volume} {21}},\ \bibinfo {pages}
  {1349} (\bibinfo {year} {2010})}\BibitemShut {NoStop}%
\bibitem [{\citenamefont {Cavagna}\ \emph {et~al.}(2005)\citenamefont
  {Cavagna}, \citenamefont {Castello}, \citenamefont {Giardina}, \citenamefont
  {Grigera}, \citenamefont {Jelic}, \citenamefont {Melillo}, \citenamefont
  {Mora}, \citenamefont {Parisi}, \citenamefont {Silvestri},\ and\
  \citenamefont {Walczak}}]{cavagna}%
  \BibitemOpen
  \bibfield  {author} {\bibinfo {author} {\bibfnamefont {A.}~\bibnamefont
  {Cavagna}}, \bibinfo {author} {\bibfnamefont {L.~D.}\ \bibnamefont
  {Castello}}, \bibinfo {author} {\bibfnamefont {I.}~\bibnamefont {Giardina}},
  \bibinfo {author} {\bibfnamefont {T.}~\bibnamefont {Grigera}}, \bibinfo
  {author} {\bibfnamefont {A.}~\bibnamefont {Jelic}}, \bibinfo {author}
  {\bibfnamefont {S.}~\bibnamefont {Melillo}}, \bibinfo {author} {\bibfnamefont
  {T.}~\bibnamefont {Mora}}, \bibinfo {author} {\bibfnamefont {L.}~\bibnamefont
  {Parisi}}, \bibinfo {author} {\bibfnamefont {E.}~\bibnamefont {Silvestri}}, \
  and\ \bibinfo {author} {\bibfnamefont {M.~V.~A.}\ \bibnamefont {Walczak}},\
  }\href@noop {} {\bibfield  {journal} {\bibinfo  {journal} {J. Stat. Phys.}\
  }\textbf {\bibinfo {volume} {158}},\ \bibinfo {pages} {601} (\bibinfo {year}
  {2005})}\BibitemShut {NoStop}%
\bibitem [{\citenamefont {Pearce}\ \emph {et~al.}(2014)\citenamefont {Pearce},
  \citenamefont {Miller}, \citenamefont {Rowlands},\ and\ \citenamefont
  {Turner}}]{pearce2014role}%
  \BibitemOpen
  \bibfield  {author} {\bibinfo {author} {\bibfnamefont {D.~J.}\ \bibnamefont
  {Pearce}}, \bibinfo {author} {\bibfnamefont {A.~M.}\ \bibnamefont {Miller}},
  \bibinfo {author} {\bibfnamefont {G.}~\bibnamefont {Rowlands}}, \ and\
  \bibinfo {author} {\bibfnamefont {M.~S.}\ \bibnamefont {Turner}},\
  }\href@noop {} {\bibfield  {journal} {\bibinfo  {journal} {Proceedings of the
  National Academy of Sciences}\ }\textbf {\bibinfo {volume} {111}},\ \bibinfo
  {pages} {10422} (\bibinfo {year} {2014})}\BibitemShut {NoStop}%
\bibitem [{\citenamefont {Barberis}\ and\ \citenamefont
  {Peruani}(2016)}]{barberis}%
  \BibitemOpen
  \bibfield  {author} {\bibinfo {author} {\bibfnamefont {L.}~\bibnamefont
  {Barberis}}\ and\ \bibinfo {author} {\bibfnamefont {F.}~\bibnamefont
  {Peruani}},\ }\href@noop {} {\bibfield  {journal} {\bibinfo  {journal} {Phys\
  Rev\ Lett}\ }\textbf {\bibinfo {volume} {117}},\ \bibinfo {pages} {248001}
  (\bibinfo {year} {2016})}\BibitemShut {NoStop}%
\bibitem [{\citenamefont {Panait}\ and\ \citenamefont {Luke}(2005)}]{panait}%
  \BibitemOpen
  \bibfield  {author} {\bibinfo {author} {\bibfnamefont {L.}~\bibnamefont
  {Panait}}\ and\ \bibinfo {author} {\bibfnamefont {S.}~\bibnamefont {Luke}},\
  }\href@noop {} {\bibfield  {journal} {\bibinfo  {journal} {Autonomous Agents
  and Multi-Agent Systems}\ }\textbf {\bibinfo {volume} {11}},\ \bibinfo
  {pages} {387} (\bibinfo {year} {2005})}\BibitemShut {NoStop}%
\bibitem [{\citenamefont {Bu{\c s}oniu}\ \emph {et~al.}(2008)\citenamefont
  {Bu{\c s}oniu}, \citenamefont {Bab{\v u}ska},\ and\ \citenamefont
  {Schutter}}]{busoniu}%
  \BibitemOpen
  \bibfield  {author} {\bibinfo {author} {\bibfnamefont {L.}~\bibnamefont
  {Bu{\c s}oniu}}, \bibinfo {author} {\bibfnamefont {R.}~\bibnamefont {Bab{\v
  u}ska}}, \ and\ \bibinfo {author} {\bibfnamefont {B.}~\bibnamefont
  {Schutter}},\ }\href@noop {} {\bibfield  {journal} {\bibinfo  {journal} {IEEE
  transactions on systems, man, and cybernetics-part C: application and
  reviews}\ }\textbf {\bibinfo {volume} {38}} (\bibinfo {year}
  {2008})}\BibitemShut {NoStop}%
\bibitem [{\citenamefont {Milinski}\ and\ \citenamefont
  {Heller}(1978)}]{milinski}%
  \BibitemOpen
  \bibfield  {author} {\bibinfo {author} {\bibfnamefont {M.}~\bibnamefont
  {Milinski}}\ and\ \bibinfo {author} {\bibfnamefont {R.}~\bibnamefont
  {Heller}},\ }\href@noop {} {\bibfield  {journal} {\bibinfo  {journal}
  {Nature}\ }\textbf {\bibinfo {volume} {275}},\ \bibinfo {pages} {642}
  (\bibinfo {year} {1978})}\BibitemShut {NoStop}%
\bibitem [{\citenamefont {Pitcher}\ \emph {et~al.}(1982)\citenamefont
  {Pitcher}, \citenamefont {Magurran},\ and\ \citenamefont
  {Winfield}}]{pitcher}%
  \BibitemOpen
  \bibfield  {author} {\bibinfo {author} {\bibfnamefont {T.}~\bibnamefont
  {Pitcher}}, \bibinfo {author} {\bibfnamefont {A.}~\bibnamefont {Magurran}}, \
  and\ \bibinfo {author} {\bibfnamefont {I.}~\bibnamefont {Winfield}},\
  }\href@noop {} {\bibfield  {journal} {\bibinfo  {journal} {Behav Ecol
  Sociobiol}\ }\textbf {\bibinfo {volume} {10}},\ \bibinfo {pages} {149}
  (\bibinfo {year} {1982})}\BibitemShut {NoStop}%
\bibitem [{\citenamefont {Sutton}\ and\ \citenamefont {Barto}(1998)}]{sutton}%
  \BibitemOpen
  \bibfield  {author} {\bibinfo {author} {\bibfnamefont {R.}~\bibnamefont
  {Sutton}}\ and\ \bibinfo {author} {\bibfnamefont {A.}~\bibnamefont {Barto}},\
  }\href@noop {} {\emph {\bibinfo {title} {Reinforcement Learning: An
  Introduction}}}\ (\bibinfo  {publisher} {MIT Press},\ \bibinfo {address}
  {Cambridge, MA},\ \bibinfo {year} {1998})\BibitemShut {NoStop}%
\bibitem [{\citenamefont {Watkins}(1992)}]{watkins}%
  \BibitemOpen
  \bibfield  {author} {\bibinfo {author} {\bibfnamefont {C.}~\bibnamefont
  {Watkins}},\ }\href@noop {} {\bibfield  {journal} {\bibinfo  {journal}
  {Machine Learning}\ }\textbf {\bibinfo {volume} {8}},\ \bibinfo {pages} {279}
  (\bibinfo {year} {1992})}\BibitemShut {NoStop}%
\end{thebibliography}%


%

\end{document}